# Towards Quranic reader controlled by speech

Yacine Yekache, Yekhlef Mekelleche, Belkacem Kouninef
Institut National des Télécommunications et des TIC
BP 1518 Oran El Mnaouer 31000 Oran, ALGERIA

*Abstract*—In this paper we describe the process of designing a task-oriented continuous speech recognition system for Arabic, based on CMU Sphinx4, to be used in the voice interface of Quranic reader.
The concept of the Quranic reader controlled by speech is presented, the collection of the corpus and creation of acoustic model are described in detail taking into account a specificities of Arabic language and the desired application.

## I. INTRODUCTION

Automatic speech recognition (ASR) is the technology that permits the communication with machine using speech. There are several applications that use this technology such as hands-free operation and control as in cars or person with disabilities, automatic dictation, government information systems, automatic query answering, telephone communication with information systems, etc.

The most dominant approach for ASR system is the statistical approach Hidden Markov Model(HMM), trained on corpora that contain speech resource from a large number of speakers to achieve acceptable performance; unfortunately there is a lack of this corpus for Arabic language. In this work we collected a new corpus called Quranic reader command and control which we will use to create an acoustic model using "sphinx train"

## II. QURANIC READER AND ARABIC LANGUAGE

### A. Quranic reader

Quran is the central religious text of Islam, which is the verbatim word of God and the Final Testament, following the Old and New Testaments. It is regarded widely as the finest piece of literature in the Arabic language. The Quran consists of 114 chapters of varying lengths, each known as a sura. Chapters are classed as Meccan or Medinan, depending on when (before or after Hijra) the verses were revealed. Chapter titles are derived from a name or quality discussed in the text, or from the first letters or words of the sura.

There is a crosscutting division into 30 parts of roughly equal division, ajza, each containing two units called ahzab, each of which is divided into four parts (rub 'al-ahzab).

The Quran is the muslims way of life and the guidance from Allah for that every muslim should read, listen and memorize it; nowadays there are computer tools used for this purpose, the interaction with this tools is by using a mouse or a keyboard but in some situation it is difficult to use them for example when driving a car or for blind person; so our goal is to create a Quranic reader controlled by speech

### B. Arabic language

Quran is revealed in Arabic for that it is the official language of 23 countries and has many different, geographically distributed spoken varieties, some of which are mutually unintelligible. Modern Standard Arabic (MSA) is widely taught in schools, universities, and used in workplaces, government and the media.

Standard Arabic has basically 34 phonemes, of which six are vowels, and 28 are consonants. A phoneme is the smallest element of speech units that makes a difference in the meaning of a word, or a sentence. The correspondence between writing and pronunciation in MSA falls somewhere between that of languages such as Spanish and Finnish, which have an almost one-to-one mapping between letters and sounds, and languages such as English and French, which exhibit a more complex letter-to-sound mapping[1] .

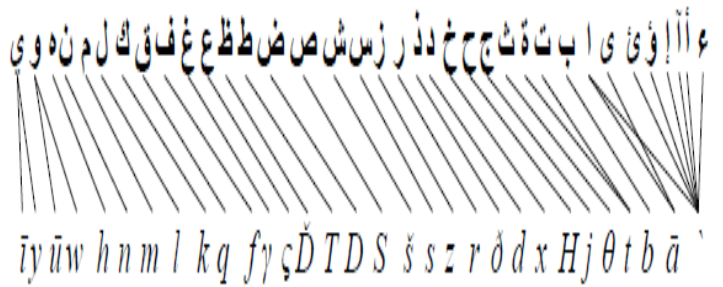

Figure 1. letter to sound mapping in Arabic

$$\widehat{W} = \underset{W}{argmax}\, P\left(W/X\right) \qquad (1)$$

## III. STATISTICAL APPROACH FOR SPEECH RECOGNITION

### A. Mathematical formulation of ASR problem

A block diagram of this fundamental approach to speech recognition is given in Fig.2, which shows a sentence W being converted to a speech signal s[n] via the speech production process. The speech signal is then spectrally analyzed (by the acoustic processor) giving the spectral representation, X = (X1, X2, . . . , XL) for the L frames of the speech signal. The linguistic decoder solves for the maximum likelihood sentence that best matches X (i. e., maximizes the probability of W given X) via the Bayesian formulation:

$$\widehat{W} = \underset{W}{argmax} \frac{P(W/X)P(W)}{P(X)} \quad (2)$$

$$\widehat{W} = \underset{W}{argmax} \underbrace{P_A(X/W)}_{Step\ 3} \underbrace{P_L(W)}_{Step\ 1\ Step\ 2} \quad (3)$$

The maximization of (1) is converted to (2) using the Bayes rule, and since the denominator term P(X) is independent of W, it can be removed leading to the three-step solution of (3). Here we explicitly denote the acoustic model by labeling P(X|W) as PA(X|W), where A denotes the set of acoustic models of the speech units used in the recognizer, and we denote P(W) as PL (W) for the language model describing the probabilities of various word combinations. The process of determining the maximum-likelihood solution is to first train (offline) the set of acoustic models so that step 1 in (3) can be evaluated for each speech utterance. The next step is to train the language model for step 2, so that the probability of every word sequence that forms a valid sentence in the language model can be evaluated. Finally step 3 is the heart of the computation, namely a search through all possible combinations of words in the language model to determine the word [2]

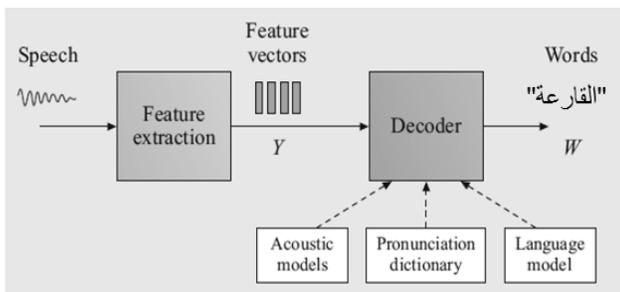

Figure 2. Architecture of an HMM based recognizer

### B. CMU Sphinx

Sphinx4 is a software implementation of HMM speech recognizer, it's architecture is highly flexible, Each labelled element in Figure (3) represents a module that can be easily replaced, allowing researchers to experiment with different module implementations without needing to modify other portions of the system. The main blocks in Sphinx-4 architecture are frontend, decoder and Linguist [3]

**Front End**: it parameterizes an input signal into a sequence of output features. It performs Digital Signal Processing on the incoming data.

--Feature: The outputs of the front end are features, used for decoding in the rest of the system.

**Linguist**: Or knowledge base, it provides the information the decoder needs to do its job, this sub-system is where most of the adjustments will be made in order to support Arabic recognition it is made up of three modules which are:

--Acoustic Model: Contains a representation of a sound, created by training using many acoustic data.

--Dictionary: It is responsible for determining how a word is pronounced.

--Language Model: It contains a representation (often statistical) of the probability of occurrence of words.

**Search Graph**: The graph structure produced by the linguist according to certain criteria (e.g., the grammar), using knowledge from the dictionary, the acoustic model, and the language model.

**Decoder**: It reads features from the front end, couples this with data from the knowledge base and feedback from the application, and performs a search to determine the most likely sequences of words that could be represented by a series of features.

**The Configuration Manager**: This sub-system is designed in order to make Sphinx-4 pluggable and flexible. Configuration information is provided in XML files, and the actual code uses the Configuration Manager to look up the different modules and sub-systems.

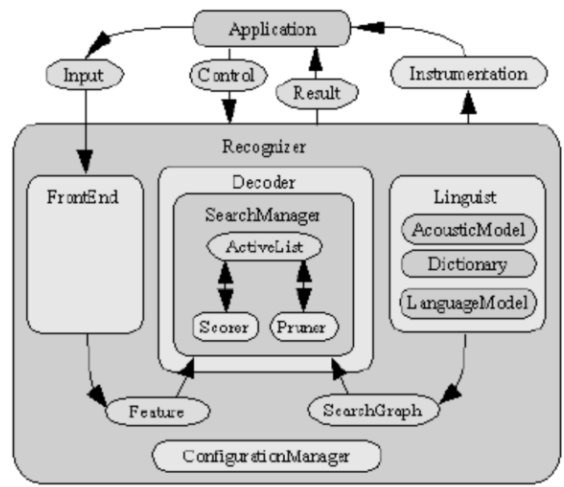

Figure 3. Sphinx-4 architecture

The Sphinx framework was originally designed for English, but nowadays it supports also, among others, Spanish, French and Mandarin. However, there are limited acoustic model available for Arabic [4, 5, 6].

Our goal is to use Sphinx to build Quranic reader controlled by speech ; the main tuning that we should do in the frame work is the creation of the acoustic model and the application that interact with the framework. In the following section we will describe the creation of the acoustic model.

## IV. ACOUSTIC MODELING

The creation of acoustic model pass through two steps, the first is data collection (corpus) and the second is training the model using this data,

### A. Data collection (corpus)

The following elements were required to train a new acoustic model:
• Audio data with recorded speech;
• Transcription of each audio file;
• Dictionary with phonetic representations of all words appearing in the transcriptions;
• List of phonemes (and sounds) appearing in the transcriptions

### B. Speech collection

We prepared a text file which contain 114 suras name's, famous receiters names and control words that must be correctly recognized for the Quranic reader to function properly, The amount of audio data required to properly train the model depends on the type of the model. For a simple command-and-control one-speaker system the amount of data can be fairly low. For multi-speaker systems the amount of required audio increases and this is our case.

After selecting text for the recognizer, recording of this chosen data is required. For this work, the recording has been taken place using 50 speakers from different region of Algeria; their personal profile includes information like name, gender, age, and other information like Environmental condition of recording (for example: class room condition, sources of noise like fan, generator sound etc) Technical details of device (pc, microphone specification)

The audio file was recorded using sampling rate of 16KHZ and 16 bit per sample, this rate was chosen because it provide more accurate high frequency information, after that the splitting by command and word was done manually and saved in .wav format[7]. Each file has been named using this convention: speakername-commandID.wav for example a file with the name yacine-001 mean that this file is recorded by yacine and it contain the Fatiha word

These audio files was divided into two sets, the first is composed of data from 20 males and 15 female used to train the acoustic model and the second composed of data from 10 males and 5 female for testing purpose.

### C. Transcription file

The second step is the transcription of the training set of the collected audio files; any error in the transcription will mislead the training process later. The transcription process is done manually, that is, we listen to the recording then we match exactly what we hear into text even the silence or the noise should be represented in the transcription.

### D. Phonetic dictionary

In this step we mapped each word in the vocabulary to a sequence of sound units representing pronunciation; that it contained all words with all possible variants of their pronunciation, to take into account pronunciation variability, caused by various speaking manners and the specificity of Arabic. . Careful preparation of phonetic dictionary prevents from incorrect association of a phoneme with audio parameters of a different phoneme which would effect in decreasing the model's accuracy.

For example

صِفْر صِ ف ر
خمسة خَ م س ه
النَّاس أنْ ن ا س

### E. List of phoneme

This is a file which contain all the acoustic units that we want to train model for, The SPHINX-4 does not permit us to have units other than those in our dictionaries. All units in the dictionary must be listed here. In other words, phone list must have exactly the same units used in your dictionaries, no more and no less. The file has one phone in each line, no duplicity is allowed.

TABLE1. WORDS USED IN THE CORPUS

| Sura name | | |
|---|---|---|
| | الفاتحة | 001 |
| | البقرة | 002 |
| | آل عمران | 003 |
| | النساء | 004 |
| | ......... | 114 |
| Reciters names | الغامدي | 115 |
| | السديسي | 116 |
| | العجمي | 117 |
| | الحذيفي | 118 |
| Control | تلاوة | 119 |
| | إنهاء | 120 |
| | توقف | 121 |
| | إستمر | 122 |
| | تكرار | 123 |
| | تحفيظ | 124 |
| | تفسير | 125 |
| | إنتقل | 126 |
| | بحث | 127 |
| | سورة | 128 |
| | آية | 129 |
| | حزب | 130 |
| | جزء | 131 |
| | نفد | 132 |
| Arabic digit | صفر | 133 |
| | واحد | 134 |
| | .. | .... |
| | تسعة | 142 |

### F. Acoustic model training

Before acoustic modeling we should extract features vectors from the speech for a purpose of training. The dominating feature extraction technique known as Mel-Frequency Cepstral Coefficients (MFCC) was applied to extract features from the set of spoken utterances. A feature vector Y represents unique characteristics of each recorded utterance,

The most widely used method of building acoustic models is HMMs, Each base phone q is represented by a continuous density HMM of the form illustrated in Fig.(4) with transition parameters {aij } and output observation distributions {bj ()}.entry and exit states are nonemitting and they are included to simplify the process of concatenating phone models to make words.

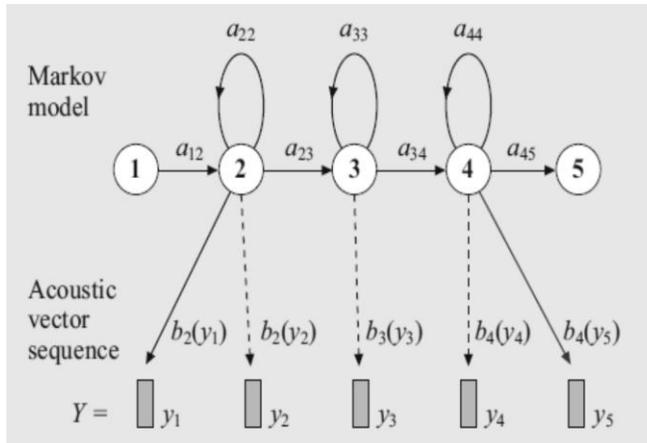

Figure 4.   HMM based phone model

Each word in our corpus should be modeled using HMM,the parameter aij and bj are estimated from our collected corpus using expectation maximization (EM). For each utterance, the sequence of base forms is found and the corresponding composite HMM constructed.

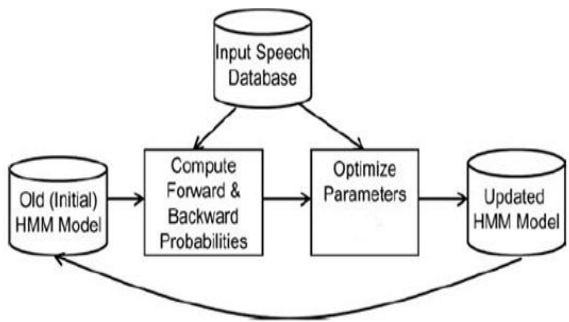

Figure 5.   The Baum-Welch training procedure based on a given training set of utterances

A forward–backward alignment is used to compute state occupation probabilities (the E step) and the means and covariances are then estimated via simple weighted averages(theMstep) [8]

To create the acoustic model we used sphinx train, which need as input the recorded speech, transcription, dictionary and phoneme files to produce the acoustic model. Much of Sphinx Train's configuration and setup uses scripts written in the programming language Perl.

V.   CONCLUSION

In this paper we reported the first steps toward developing Quranic reader controlled by speech using sphinx4 framework, in this steps we specified the words that should be recognized and collected a corpus used to train the acoustic model with sphinx train.

Further we will integrate the acoustic model in sphinx 4 and build an application that interacts with sphinx framework.